\input amstex
\magnification 1200
\documentstyle{amsppt}

\define\a{\alpha}  \define\aw{{\widetilde\alpha}}
 \define\ah{{\widehat\alpha}}
\predefine\overlineunder{\b} \redefine\b{\beta} 
  
\predefine\cedilla{\c} \redefine\c{\psi} 
  
\redefine\d{\delta}

 \define\5d{{\dot\varepsilon}} 
\define\f{\epsilon}  
 
\define\g{\gamma}

\redefine\l{\lambda}  
 
\define\m{\mu}  \define\mw{{\widetilde\mu}}
 
\define\n{\nu}

\define\q{\phi}  
 
\define\r{\rho}  \define\rw{{\widetilde\rho}}
 
\define\s{\sigma}  \define\sw{{\widetilde\sigma}}
 
\redefine\t{\theta}

\predefine\wP{\wp} \redefine\wp{{\dot\varpi}}

\define\y{\tau}

 \define\xW{{\widetilde x}} 
 \define\nW{{\widetilde n}} \define\mW{{\widetilde m}}

  \define\C{\Psi}
  
 \predefine\dotaccent{\D} \redefine\D{\Delta}
  
 \define\G{\Gamma} 
  
\redefine\L{\Lambda}  
  \predefine\or{\O} \redefine\O{\Omega}
  
 \predefine\paragraph{\P} \redefine\P{\varPi}
  
 \define\Q{\Phi} 
  
\predefine\section{\S} \redefine\S{\Sigma} \predefine\Sub{\Sb}
\redefine\Sb{{\overline\Sigma}}

 \define\1{{\dot 1}} \define\2{{\dot 2}}
\define\8{\infty} \define\0{\emptyset} 
\define\pa{\partial}

\define\({\left(} \define\){\right)} \define\[{\left[} \define\]{\right]}
   
 \define\sq{\sqrt} 
  \define\={x^\n{}\!,_\aw \[\pa_\m,\pa_\n\]x^\aw=0}
\define\I{\int} \define\X{\times}

\loadbold \TagsOnRight \NoRunningHeads \pageno=1 \topmatter
\title
Unified Field Theory from Enlarged Transformation Group.  The Consistent Hamiltonian
\endtitle
\author
Dave Pandres, Jr. and Edward L. Green\footnotemark
\endauthor
\endtopmatter
\document
\flushpar ( Pandres, D. and Green, E., Unified field theory from enlarged
transformation group. The consistent Hamiltonian., {\it International Journal of
Theoretical Physics}, {\bf 42}, 1849-1873. )

\phantom{D}

 \flushpar $^1$Department of Mathematics and Computer Science, North Georgia
College\linebreak \hphantom{1}and State University, Dahlonega, Georgia 30597,
(706)864-1809,  egreen\@ngcsu.edu

\subhead Abstract
\endsubhead
\flushpar A theory has been presented previously in which the
geometrical structure of a real four-dimensional space time
manifold is expressed by a real orthonormal tetrad, and the group
of diffeomorphisms is replaced by a larger group. The group
enlargement was accomplished by including those transformations to
anholonomic coordinates under which conservation laws are
covariant statements. Field equations have been obtained from a
variational principle which is invariant under the larger group.
These field equations imply the validity of the Einstein equations
of general relativity with a stress-energy tensor that is just
what one expects for the electroweak field and associated
currents. In this paper, as a first step toward quantization, a
consistent Hamiltonian for the theory is obtained.  Some
concluding remarks are given concerning the need for further
development of the theory. These remarks include discussion of a
possible method for extending the theory to include the strong
interaction.

\vskip .5in

\subhead
1. INTRODUCTION
\endsubhead
In Sections 1 and 2, we describe a theory in which the classical (unquantized)
gravitational and electroweak fields appear as manifestations of geometrical structure
in a real four-dimensional space-time manifold. In Section~3, we obtain the Hamiltonian
for the theory as a first step toward quantizing the theory. In Section~4, we make some
concluding remarks concerning the further development of the theory. One of these
remarks suggests a method for extending the theory to include the strong interaction.
[NOTE: In several prior papers, one of us (Pandres, 1981, 1984A, 1984B, 1995, 1998,
1999), has based the theory, not on a manifold, but on a space in which {\sl paths,}
rather than points are the primary elements. In this paper, however, we show that the
theory can be based entirely on a manifold.

It is well known that any general relativistic metric $g_{\m\n}$ may be expressed in terms of an orthonormal tetrad of vectors $h^i{}\!_\m$. The expression is
$$
g_{\m\n} = g_{ij} h^i{}\!_\m h^j{}\!_\n    \tag1
$$
where $g_{ij}= g^{ij}=\text{diag}(-1,1,1,1)$, and the summation convention has been
\newline adopted.
 Indices take the values $0,1,2,3$, and
$\dsize
g^{\m\n}
$
is
defined by
$g^{\m\n} g_{\n\a}= \d^\m_\a$,
where
$
\d^\m_\a
$
is the Kronecker delta. Latin (tetrad) indices are raised and lowered by using $g^{ij}$ and $g_{ij}$, just as Greek (space time) indices are raised and lowered by using $g^{\m\n}$ and $g_{\m\n}$. Partial differentiation is denoted by a comma. Covariant differentiation with respect to the Christoffel symbol
$\dsize
\G^\a{}\!_{\m\n}
= {\tsize \frac{1}{2}}g^{\a\s}\(g_{\s\m},_\n + g_{\s\n},_\m - g_{\m\n},_\s\)
$
is denoted by a semicolon.
\subhead
1.1. Motivation
\endsubhead
We recall (Pandres 1962, 1999) an argument which is a generalization of the ``elevator'' argument that led Einstein from special relativity to general relativity. The special relativistic equation of motion for a free particle is
$$
\frac{d^2 x^i}{ds^2}=0  \quad , \tag2
$$
where $-ds^2=g_{ij}dx^i dx^j$.
Consider the image-equation of this free-particle equation under the
transformation
$$
dx^i = h^i{}\!_\m dx^\m     \tag3
$$
where the curl
$
f^i{}\!_{\m\n}=h^i{}\!_\n,_\m - h^i{}\!_\m,_\n
$
is not zero. Eq.~(3) establishes a one-to-one correspondence between {\sl coordinate increments} \/ $dx^i$ and $dx^\m$. Since
$
h^i{}\!_\n,_\m - h^i{}\!_\m,_\n
$
is not zero, we cannot integrate Eq.~(3) to get a one-to-one correspondence between coordinates
$x^i$ and $x^\m$. However, it follows from Eq.~(3) that
$\dsize
\frac{dx^i}{ds} = h^i{}\!_\m \frac{dx^\m}{ds}
$ .
Upon differentiating this with respect to $s$, using the chain rule, and multiplying by
$h_i{}^\a$, we see that Eq.~(2) may be written
$$
\frac{d^2 x^\a}{ds^2} +  h_i{}^\a h^i{}\!_{\m,\n} \frac{dx^\m}{ds} \frac{dx^\n}{ds}= 0 \quad .
\tag4
$$
We follow Eisenhart (1925) in defining
Ricci rotation coefficients by $\g^i{}\!_{\m\n} = h^i\!_{\m ;\n}= h^i{}\!_{\m,\n} - h^i{}\!_\s \G^\s{}\!_{\m\n}.$
Multiplication by $h_i{}^\a$ gives $h_i{}^\a h^i{}\!_{\m,\n} = \G^\a{}\!_{\m\n} +     \g^\a{}\!_{\m\n}$, and upon using this in Eq.~(4) we have

$$
\frac{d^2 x^\a}{ds^2} +  \G^\a{}\!_{\m\n} \frac{dx^\m}{ds} \frac{dx^\n}{ds}=
-\g^\a{}\!_{\m\n} \frac{dx^\m}{ds} \frac{dx^\n}{ds}
\quad .\tag5
$$
The relation
$
\g_{\m\n i}=   h^j{}\!_\m \g_{j\n\a} h_i{}\!^\a
$
illustrates our general method for converting between Greek and Latin indices.

Now, the affine connection for spin in general relativity is expressed in terms of the
Ricci rotation coefficients by\;
$
\G_\m = {\tsize \frac{1}{8}}
\g_{ij\m} \(\g^i \g^j - \g^j \g^i\) + a_\m I \; ,
$
where the $\g^i$ are the Dirac matrices of special relativity, $I$ is the
identity matrix, and $a_\m$ is an arbitrary vector. It is well known that the spin connection contains complete information about the electromagnetic field, and that one half of Maxwell's equations are identically satisfied on account of the existence of the spin connection.
Furthermore, the manner in which the electromagnetic field enters the spin
connection is in agreement with the principle of minimal electromagnetic
coupling. An understanding of the spinor calculus in Riemann space, and the
role played by the spin connection, was gained through the work of many
investigators during the decade after Dirac's discovery of the relativistic
theory of the electron; see, e.g., Bade and Jehle (1953) for a general review.
Many of these investigators recognized the description of the electromagnetic
field as part of the spin connection. An especially lucid discussion of this
has been given by Loos (1963). The subsequent unification of the
electromagnetic and weak fields by Weinberg (1967), and Salam (1968) causes us
to expect that the spin connection might also contain a description of the
weak field.

We now recall (Pandres, 1995) calculations that suggest that the electroweak field is
described by $M_{\m\n i}$, the ``mixed symmetry'' part of $\g_{\m\n i}$ under the
permutation group on three symbols. One may object to using $\g_{\m\n i}$ to describe
the electoweak field since $\g_{i j \m}$ is used in the spin connection. However, these
geometric objects cannot be considered to be the same since the method of converting
from one to the other is not a diffeomorphism.  The method for converting between Greek
and Latin indices involves $h^i{}\!_\m$.  Thus the components of $\g_{\m\n i}$ are
quite independent of the components of $\g_{i j \m}$, although if one is zero the other
is also zero.  The totally symmetric part of $\g_{\m\n i}$ vanishes because it is
antisymmetric in $\m$ and $\n$. Thus, we have $\dsize \g_{\m\n i}=M_{\m\n i}+A_{\m\n i}
, $ where $A_{\m\n i}$ is the totally antisymmetric part. Clearly, $A^\a{}\!_{\m\n}$
makes no contribution to the right side of Eq.~(5), so
$$
\frac{d^2 x^\a}{ds^2} +  \G^\a{}\!_{\m\n} \frac{dx^\m}{ds} \frac{dx^\n}{ds}=
\frac{dx^\m}{ds} M_\m{}\!^\a{}\!_i \, v^i
\quad ,\tag6
$$
where
$\dsize
v^i=\frac{dx^i}{ds}
$ is the (constant) first integral of Eq.~(2).
The totally antisymmetric part of $\g_{\m\n i}$ is
$$
A_{\m\n i}={\tsize\frac{1}{3}} \(\g_{\m\n i} + \g_{i\m\n}+\g_{\n i\m}\)
\;  . \tag7
$$
Thus, the mixed symmetry part is  $M_{\m\n i}=\g_{\m\n i}-A_{\m\n i}$,
so, we have
$$
M_{\m\n i}={\tsize\frac{1}{3}}\(2\g_{\m\n i}
                        -\g_{i\m\n} -\g_{\n i\m}\) \;   .\tag8
$$
The antisymmetry of $\g_{\m\n i}$ in its first two indices may be used to obtain an
expression for $M_{\m\n i}$ in terms of $f_{i\m\n}$. We have
$
f_{i\m\n}= h_{i\n},_\m - h_{i\m},_\n
   =h_{i\n ;\m} - h_{i\m ;\n}
$ ,
so that
$
f_{i\m\n} = \g_{i \n\m} - \g_{i \m\n}
$.
If we subtract from this the corresponding expressions for $f_{\m\n i}$ and
$f_{\n i\m}$, we see that
$
\g_{\m\n i} = \tsize{\frac{1}{2}} \(f_{i\m\n} - f_{\m\n i} - f_{\n i\m}\)
$.
By using this and the corresponding expressions for $\g_{i\m\n}$ and $\g_{\n i\m}$
in Eq.~(8), we obtain
$$
M_{\m\n i}={\tsize\frac{1}{3}} \(2f_{i\m\n} - f_{\m\n i} - f_{\n i\m}\)\quad ,  \tag9
$$
which may be written
$$
M_{\m\n i}={\tsize\frac{1}{3}}\(2 \d^n_i  \d^\a_\m \d^\s_\n
                      - h^n{}\!_\m \d^\a_\n h_i{}\!^\s
                      - h^n{}\!_\n  h_i{}\!^\a  \d^\s_\m\)f_{n\a\s} \; ,\tag10
$$
where $\d^\a_\m$ is the Kronecker delta. It is important to notice that Eq.~(10) may be rewritten into the form
$$
M_{\m\n i}={\tsize\frac{1}{3}}\(2 \d^n_i \d^\a_\m \d^\s_\n
                      - h^n{}\!_\m \d^\a_\n h_i{}\!^\s
                      - h^n{}\!_\n  h_i{}\!^\a  \d^\s_\m\)
                        \frak F_{n\a\s} \quad  ,\tag11
$$
where
$$
\frak F_{i\m\n} = f_{i\m\n} + e_{0ijk} h^j{}\!_\m h^k{}\!_\n \quad , \tag12
$$
and $e_{nijk}$ is the Levi-Civita symbol.
In rewriting Eq.~(10) as Eq.~(11), we have used the easily verifiable fact that
$$
\(2 \d^n_i \d^\a_\m \d^\s_\n
                      - h^n{}\!_\m \d^\a_\n h_i{}\!^\s
                      - h^n{}\!_\n  h_i{}\!^\a  \d^\s_\m\)
                         e_{0njk} h^j{}\!_\a h^k{}\!_\s = 0 \quad .
$$

Now, $\frak F_{i\m\n}$ is the usual field strength (see, e.g., Nakahara, 1990) for a
$U(1)\X SU(2)$ gauge field, {\sl provided that $h^i{}\!_\m$ is transformed on its
tetrad indices as a gauge potential,} rather than as a Lorentz vector.  We wish to make
it clear that we will not require that $h^i{}\!_\m$ be transformed as a gauge
potential.  In our view, the need for such a transformation rule arises from the fact
that coordinate transformations are limited to the diffeomorphisms. In Section 2, we
enlarge the group of diffeomorphisms to the conservation group. The mass-changing
effect of a non-diffeomorphic conservative transformation is similar to what one would
get if $h^i{}\!_\m$ were to be transformed as a gauge potential.  It is eminently
reasonable that when a particle is subjected to a rotation in isospace the
gravitational field may change.

From Eq.~(11), we see that in the expression, Eq.~(10), for $M_{\m\n i}$, the curl
$f_{n\a\s}$ may simply be replaced by the gauge field $\frak F_{n\a\s}$. The $ \frak
F_{n\a\s} $ may be viewed as a field with ``bare'' or massless quanta, which are
``clothed'' by the factor \newline $ {\tsize\frac{1}{3}}\(2 \d^n_i \d^\a_\m \d^\s_\n
                      - h^n{}\!_\m \d^\a_\n h_i{}\!^\s
                      - h^n{}\!_\n  h_i{}\!^\a  \d^\s_\m\)
$ , and thus may acquire mass. The analysis in Section 2.4 suggests that $M_{\m\n i}$
may describe the physical electroweak field as it appears in the appropriate way in our
Lagrangian, and in the stress-energy tensor of the Einstein equations. For this
identification to be valid, the quantity $\dsize M_{\m\n 0}={\tsize\frac{1}{3}}
\(2f_{0\m\n} - f_{\m\n 0} - f_{\n 0\m}\) $ must describe the electromagnetic field;
hence, it must be the curl of a vector. The presence of the terms $- f_{\m\n 0} - f_{\n
0\m}$ may cause one to ask how $M_{\m\n i}$ can be identified as the electroweak field.

Our answer is this: The orthodox physical interpretation, which we adopt, is that $h^i{}\!_\m$
describes an observer-frame.  Now, {\sl if $h^i{}\!_\m$ describes a freely falling,
nonrotating observer frame, our expression for $M_{\m\n 0}$ reduces to
$M_{\m\n 0}={\tsize\frac{1}{3}} f_{0\m\n}$.} This may be seen as follows.
The condition for a freely falling, nonrotating frame (Synge, 1960) is
$h_{i\n;\a} h_0{}\!^\a =0$. In terms of the Ricci rotation
coefficients, the condition is \;
$\dsize
\g_{\m\n 0}=0 .
$ \;
From this and Eq.~(8), we see that for an $h^i{}\!_\m$ which describes a
freely falling, nonrotating observer frame,
$\dsize
M_{\m\n 0}={\tsize\frac{1}{3}}\(\g_{0\n\m} -\g_{0\m\n}\)
          ={\tsize\frac{1}{3}}\(h_{0\n;\m} - h_{0\m;\n}\)
          ={\tsize\frac{1}{3}}\(h_{0\n},\!_\m - h_{0\m},\!_\n\)
          ={\tsize\frac{1}{3}} f_{0\m\n}
$.
Moreover, in the nonrelativistic limit (i.e., for $v^1, v^2, v^2$ small compared to one),
the electromagnetic term
$\dsize
\frac{dx^\m}{ds} M_\m{}\!^\a{}\!_0 \, v^0
$
dominates the right side of Eq.~(6).

\vskip 0.5in
\subhead 2. GRAVITATIONAL AND ELECTROWEAK UNIFICATION
\endsubhead

It is clear that no meaningful physics can be done without an observer. Thus the principle of parsimony (Occam's razor) suggests that we consider a theory in which the observer-frame
$h^i{}\!_\m$ is the {\sl only} fundamental field; i.e., in which geometrical structure is expressed by $h^i{}\!_\m$, rather than by $g_{\m\n}$. For this purpose we need an invariant Lagrangian constructed from $h^i{}\!_\m$ and its derivatives, to be used in a variational principle (analogous to the Hilbert variational principle for gravitation, but with $h^i{}\!_\m$ varied rather than $g_{\m\n}$).

\subhead
2.1. Weitzenb\"ock Invariants
\endsubhead
Soon after Einstein (1928A, 1928B) introduced tetrads into physics,
Weitzenb\"ock (1928) considered quantities constructed from $h^i{}\!_\m$ and its derivatives which are invariant under the diffeomorphisms (the coordinate transformations of general relativity).
Weitzenb\"ock listed the invariants
$$
A={\tsize \frac{1}{4}}f^{i\m\n} f_{\n\m i}\quad
 B={\tsize \frac{1}{4}}f^{i\m\n} f_{i\m\n}\quad
\Q ={\tsize \frac{1}{4}}C^\n C_\n\quad
\C ={\tsize\frac{1}{2}}(C^\n{}\!,_\n + C^\m h_i{}\!^\n h^i\!{}_\m,_\n)
$$
where the vector $C_\n$ is defined by
$$
C_\n = h_i{}^\m f^i{}\!_{\m\n} \quad . \tag13
$$
As Weitzenb\"ock noted, the most general Lagrangian which yields second-order field equations
that are linear in the second derivatives of $h^i{}\!_\m$ is
$\dsize
L = aA + bB  +  \q\Q  + \c\C
$,
where the coefficients
$
a,b,\q,\c
$,
are constants.

In order to optimize clarity in our discussion, we shall use an equivalent list of invariants
$$
W_1 = f^{i\m\n} f_{\n\m i}\quad
W_2 = f^{i\m\n} f_{i\m\n}\quad
W_3 = C^\n C_\n\quad
W_4 = C^\n{}\!_{;\n}.                    \tag14
$$
That our list is equivalent to Weitzenb\"ock's is clear: $\dsize W_1 = 4A , $ $\dsize
W_2 = 4B ,\; W_3 = 4\Q $ and $\dsize W_4 = C^\n{}\!,_\n + C^\m \G^\n{}\!_{\m\n}
      =       \(C^\n{}\!,_\n + C^\m h_i{}\!^\n h^i\!{}_\m,_\n\)
      - C^\m\(h_i{}\!^\n h^i\!{}_\m,_\n -\G^\n{}\!_{\m\n}\)
      = 2\C - C^\m\(h_i{}\!^\n h^i\!{}_\m,_\n   -\G^\n{}\!_{\m\n}\)
$.
We see from the definition of  $\G^\a{}\!_{\m\n}$ and Eq.~(1) that
$\dsize
\G^\n{}\!_{\m\n} = {\tsize \frac{1}{2}}g^{\s\n} g_{\s\n},_\m = h_i{}\!^\n h^i{}\!_\n,_\m
$.
Thus, we have
$\dsize
W_4 = 2\C - C^\m h_i{}\!^\n \(h^i\!{}_\m,_\n - h^i{}\!_\n,_\m\)
    = 2\C - C^\m h_i{}\!^\n f^i{}\!_{\n\m} = 2\C - C^\m C_\m = 2\C - 4\Q.
$

Clearly, we may write the Lagrangian as
$$
L = k_1 W_1 + k_2 W_2 + k_3 W_3  + k_4 W_4  \tag15
$$
where the coefficients
$
k_1, k_2, k_3, k_4
$,
are constants.

Weitzenb\"ock recognized that if the fields to be varied are just
the components of $g_{\m\n}$, then there is essentially no freedom
of choice for the coefficients. He showed that except for a common
multiplicative constant, one must choose $ a=-2,\, b=-1,\, \q
=-4,\, \c=4; $ i.e., $ k_1= -{\tsize \frac{1}{2}} ,\; k_2=
-{\tsize \frac{1}{4}}, $ $ k_3 = 1,\; k_4 = 2 $. With this choice,
$L$ is just the Ricci scalar $R$, which is the Lagrangian for the
free gravitational field. However, since the fields to be varied
in our theory are the components of $h^i{}\!_\m$, there exists a
nondenumerable infinity of inequivalent Lagrangians corresponding
to different ratios of the constants $k_1, k_2, k_3, k_4$. Thus,
we are confronted with a dilemma that was anticipated by Einstein
(1949). He noted that with the introduction of a richer  structure
(such as our tetrad), the diffeomorphism group ``will no longer
determine the equations as strongly as in the case of the
symmetric tensor as structure.'' Einstein also suggested the
solution for this dilemma: ``Therefore it would be most beautiful,
if one were to succeed in expanding the group once more, analogous
to the step which led from special relativity to general
relativity.'' Einstein's suggestion was in accord with the
prophetic statement by Dirac (1930) that ``The growth of the use
of transformation theory, as applied first to relativity and later
to the quantum theory is the essence of the new method in
theoretical physics. Further progress lies in the direction of
making our equations invariant under wider and still wider
transformations.'' Dirac went on to remark ``This state of affairs
is very satisfactory from a philosophical point of view, as
implying an increasing recognition of the part played by the
observer in himself introducing the regularities that appear in
his observations \dots .'' Dirac's remark supports our use of the
observer-frame $h^i{}\!_\m$ as the only fundamental field.

In Sec.~2.3., we shall see that $k_1W_1 + k_2W_2$ is not invariant under a group larger
than the diffeomorphisms for any choice of the constants $k_1$ and $k_2$. By contrast,
$k_3W_3 + k_4W_4$ is invariant under a group larger than the diffeomorphisms for {\sl
arbitrary} choice of $k_3$ and $k_4$. But, $W_4 = C^\n{}\!_{;\n}$ is a covariant
divergence; so, the term $k_4 W_4$ would make no contribution to field equations.
Hence, we shall choose for our Lagrangian the invariant $W_3 = C^\n C_\n$. We shall see
that this Lagrangian is just the sum of the gravitational Lagrangian $R$ and terms
which we tentatively label as the electroweak Lagrangian $E$. The terms $R$ and $E$ are
each invariant only under the diffeomorphisms; it is their sum that is invariant under
the larger group. By using Eq.~(12), we also note that $C_\n$ may be rewritten into the
form $C_\n = h^{i\m} \frak F_{i\m\n}$. Thus, in the expression for $C_\m$, just as in
the expression for $M_{\m\n i}$, the curl $f_{i\m\n}$ may simply be replaced by the
gauge field $\frak F_{i\m\n}$.

\subhead
2.2. Holonomic and Anholonomic Coordinates
\endsubhead
It is possible to establish a one-to-one correspondence between points $x$ of the
manifold and coordinates $x^\a$ (at least in finite coordinate patches). Such
coordinates are called (Schouten, 1954) {\sl holonomic}\/ coordinates. Let
transformation coefficients $X^\aw{}\!_\m$ have a nonzero determinant, and let the
components of $X^\aw{}\!_\m$ have definite values at each point $x$. Then, these
components are one-valued functions of holonomic coordinates, i.e., $\dsize
X^\aw{}\!_\m = X^\aw{}\!_\m\(x^\s\) $. The relation
$$
dx^\aw = X^\aw{}\!_\m \(x^\s\) dx^\m  \tag16
$$
establishes a one-to-one correspondence between {\sl coordinate increments}\/ $dx^\a$ and  $dx^\aw$. The inverse relation to Eq.~(16) is
$$
dx^\m = X^\m{}\!_\aw \(x^\s\) dx^\aw  \tag17
$$
where
$X^\m{}\!_\aw\(x^\s\)$ is defined by
$\dsize
X^\m{}\!_\aw X^\aw{}\!_\n = \d^\m_\n
$.
Eq.~(16) may be integrated to give a one-to-one correspondence between coordinates $x^\m$ and $x^\aw$ if and only if
$$
X^\aw{}\!_\n,_\m - X^\aw{}\!_\m,_\n =0  \quad .  \tag18
$$
Thus, if Eq.~(18) is satisfied, the $x^\aw$ are also holonomic coordinates. If Eq.~(18) is {\sl not}\/ satisfied, then the $x^\aw$ are called (Schouten, 1954) {\sl anholonomic coordinates.}

There does not exist a one-to-one correspondence between points $x$ of the manifold and anholonomic coordinates. Thus, in an equation such as Eq.~(17), the holonomic coordinates
$x^\s$ cannot be eliminated in favor of anholonomic coordinates $x^\sw$.
A transformation to anholonomic coordinates must be accompanied by what Schouten calls a ``mitschleppen,'' i.e., a ``dragging along'' of the holonomic coordinates. (In this sense, holonomic and anholonomic coordinates are not on the same footing. They can be put on the same footing through the introduction of a path space, as we have done in several prior papers. In this paper, however, our setting is a manifold.) We can enlarge the covariance group so that it includes transformations to anholonomic coordinates, because our group elements are the transformation coefficients (which have definite values at each point $x$).

We shall need partial derivatives with respect to anholonomic as well as holonomic coordinates.
Let $F$ be a function with a definite value at each point $x$. If $x^\a$ and $x^\aw$ are both holonomic, the relation between $F,_\a$ and $F,_\aw$ is
$$
F,_\aw = F,_\m X^\m{}\!_\aw   \quad . \tag19
$$
Thus, regardless whether $x^\aw$ is holonomic or anholonomic, we may take Eq.~(19) as the
{\sl definition}\/ of $F,_\aw$ (where $x^\a$ remains holonomic). Let the coordinates $x^{\widehat{\alpha}}$
also be either holonomic or anholonomic. Then, of course,
$\dsize
F,_{\widehat{\alpha}} = F,_\m X^\m{}\!_{\widehat{\alpha}}
$,
and we easily find that
$\dsize
F,_{\widehat{\alpha}} = F,_\mw X^\mw{}\!_{\widehat{\alpha}}
$,
where
$\dsize
X^\mw{}\!_{\widehat{\alpha}} = X^\mw{}\!_\s X^\s{}\!_{\widehat{\alpha}}
$.
\subhead
2.3. The Conservation Group
\endsubhead
The transformation law for a tetrad of vectors is
$$
h^i{}\!_\m = h^i{}\!_\aw X^\aw{}\!_\m \quad . \tag20
$$
Upon differentiating Eq.~(20) with respect to $x^\n$, we have
$\dsize
h^i{}\!_\m,_\n = h^i{}\!_\aw,_\n X^\aw{}\!_\m + h^i{}\!_\aw X^\aw{}\!_\m,_\n
$\linebreak
$\dsize
= h^i{}\!_\aw,_\sw X^\sw{}\!_\n  X^\aw{}\!_\m + h^i{}\!_\aw X^\aw{}\!_\m,_\n
$. If we subtract this from the corresponding expression with $\m$ and $\n$ interchanged,
we obtain
$$
f^i{}\!_{\m\n} = f^i{}\!_{\aw\sw} X^\aw{}\!_\m X^\sw{}\!_\n
+ h^i{}\!_\aw \(X^\aw{}\!_\n,_\m - X^\aw{}\!_\m,_\n\)    \tag21
$$
where
$\dsize
f^i{}\!_{\aw\sw} = h^i{}\!_\sw,_\aw - h^i{}\!_\aw,_\sw
$. We see from Eq.~(21) that $f^i{}\!_{\m\n}$ transforms as a tensor if and only if
Eq.~(18) is satisfied, i.e., if and only if the transformation is a diffeomorphism.
We also see from Eqs.~(14) and (21) that no linear combination of $W_1$ and $W_2$ with constant coefficients is invariant under a larger group than the diffeomorphisms. By contrast,
if we multiply Eq.~(21) by
$\dsize
h_i{}^\m = h_i{}^\rw X^\m{}\!_\rw
$
and use Eq.~(13), we get
$$
C_\n = C_\aw X^\aw{}\!_\n + X^\m{}\!_\aw \(X^\aw{}\!_\n,_\m - X^\aw{}\!_\m,_\n\) \quad . \tag22
$$
We see from Eq.~(22) that $C_\n$ transforms as a vector if and only if
$$
X^\n{}\!_\aw  \(X^\aw{}\!_\n,_\m - X^\aw{}\!_\m,_\n\) = 0 \quad .   \tag23
$$
Accordingly, we recall (Pandres, 1981) that $C^\n C_\n$ is invariant under transformations that satisfy Eq.~(23).

\subsubhead
2.3.1. Conservative Coordinate Transformations
\endsubsubhead
In the discussion that led to Eq.~(23), $x^\a$ was required to be holonomic. We now relax that
requirement and allow $x^\a$ and/or $x^\aw$ to be either holonomic or anholonomic.
A transformation which satisfies Eq.~(23) is called {\sl conservative}. This terminology is
appropriate for the following reason: A relativistic conservation law is an expression of the form
$\dsize
V^\a\!,\!_\a=0
$ ,
where $V^\a$ is a vector density of weight $+1$. This is a covariant
statement under a coordinate transformation relating $x^\a$
and $x^\aw$ if and only if it implies and is implied by the relation
$\dsize
V^\aw\!,\!_\aw=0.
$
The transformation law for a vector density of weight $+1$ is
$\dsize
V^\aw
 ={\tsize\frac{\pa x}{\pa \xW}} \, X^\aw\!_\m V^\m
$ ,
where $\frac{\pa x}{\pa \xW}$ is the (non-zero) Jacobian determinant of $X^\m{}\!_\aw$.
Upon differentiating $V^\aw$ with
respect to $x^\aw$,
we obtain
$
V^\aw{}\!,_\aw
=\(\frac{\pa x}{\pa \xW} \, X^\aw\!_\m\)\!,_\aw V^\m
+ \frac{\pa x}{\pa \xW} \, V^\a{}\!,{}\!_\a .
$
For arbitrary $V^\m$, we see that a conservation law is a covariant
statement if and only if
$$
\({\tsize\frac{\pa x}{\pa \xW}} \, X^\aw\!_\m\)\!,\!_\aw = 0  \quad . \tag24
$$
For this reason, we call a coordinate transformation
{\sl conservative\/} if it satisfies Eq.~(24). Now,
$$
\({\tsize\frac{\pa x}{\pa \xW}} \, X^\aw\!_\m\)\!,\!_\aw
= \({\tsize\frac{\pa x}{\pa \xW}}\)\!,_\a X^\aw\!_\m
+ {\tsize\frac{\pa x}{\pa \xW}} X^\aw\!_\m,_\aw
=\({\tsize\frac{\pa x}{\pa \xW}}\),_\m
+ {\tsize\frac{\pa x}{\pa \xW}} X^\aw\!_\m,_\n X^\n{}\!_\aw
$$
so, if we use the well-known formula
$$
\({\tsize\frac{\pa x}{\pa \xW}}\)\!,_\m
=  {\tsize\frac{\pa x}{\pa \xW}} \, X^\aw{}\!_\n X^\n\!,\!_\aw,_\m
$$
for the derivative of a determinant, and note that
$X^\aw\!_\n  X^\n\!_\aw,_\m  =-X^\aw\!_\n,_\m X^\n\!_\aw$,
we find that Eq.~(24) is equivalent to Eq.~(23).
\subsubhead
2.3.2. The conservation group
\endsubsubhead
We now recall (Pandres, 1981) an explicit proof that the conservative
coordinate transformations form a group.
[Finkelstein (1981), however, has
pointed out that the group property follows implicitly from the derivation
given above.]
First, we note that the identity transformation $x^\aw = x^\a$ is a
conservative coordinate transformation.
Next, we consider the
result of following a coordinate transformation from $x^\a$ to $x^\aw$ by a
coordinate
transformation from $x^\aw $ to $x^\ah$. Upon differentiating
$$
X^\ah\!_\m = X^\ah\!_\rw X^\rw\!_\m  \tag25
$$
with respect to $x^\n$, subtracting the corresponding expression with $\m$ and $\n$
interchanged, and multiplying by $X^\n\!_\ah$ we obtain
$$\aligned
X^\n\!_\ah \(X^\ah{}\!_\n,_\m - X^\ah{}\!_\m,_\n\) =&
                           X^\rw\!_\m X^\sw\!_\ah \(X^\ah{}\!_\sw,_\rw- X^\ah{}\!_\rw,_\sw\)\\
            &+ X^\n\!_\rw \(X^\rw{}\!_\n,_\m - X^\rw{}\!_\m,_\n\).\endaligned\tag26
$$
We see from Eq.~(26) that if
$
X^\n\!_\rw \(X^\rw{}\!_\n,_\m - X^\rw{}\!_\m,_\n\)
$
and
$
X^\sw\!_\ah \(X^\ah{}\!_\sw,_\rw - X^\ah{}\!_\rw,_\sw\)
$
vanish, then\linebreak
$
X^\n\!_\ah \(X^\ah{}\!_\n,_\m - X^\ah{}\!_\m,_\n\)
$
vanishes. This shows that if the transformations from $x^\a$ to $x^\aw$
and from $x^\aw$ to $x^\ah$ are conservative coordinate
transformations, then the product transformation
from $x^\a$ to $x^\ah$ is a conservative coordinate transformation.
If we let $x^\ah=x^\a$, we see from
Eq.~(26) that the inverse of a conservative coordinate
transformation is a
conservative coordinate transformation.
From Eq.~(25), we see that the product of matrices
$X^\rw\!_\m$ and $X^\ah\!_\rw$ (which
represent the transformations from
$x^\a$ to $x^\aw$ and from $x^\aw$ to $x^\ah$, respectively) equals the matrix
$X^\ah\!_\m$ (which represents the product transformation from $x^\a$ to
$x^\ah$). It is obvious, and well known, that if products admit a matrix
representation in this sense, then the associative law is satisfied. This
completes the proof that the conservative coordinate
transformations form a group, which we call the conservation group.

To show that the conservation group contains the diffeomorphisms as a proper subgroup, we need only exhibit transformation coefficients which satisfy Eq.~(23), but do not satisfy Eq.~(18).
Let
$$
X^\aw\!_\n = \d^\aw_\n + \d^\aw_0 \d^2_\n x^1   \quad .        \tag27
$$
Upon differentiating Eq.~(27) with respect to $x^\m$ and subtracting the corresponding expression with $\m$ and $\n$ interchanged, we obtain
$$
X^\aw{}\!_\n,_\m - X^\aw{}\!_\m,_\n
=\d^\a_0\(\d^1_\m \d^2_\n-\d^1_\n\d^2_\m\) \quad . \tag28
$$
A nonzero component of Eq.~(28) is
$
X^{\widetilde 0} \!_2,_1 - X^{\widetilde 0} \!_1,_2 = 1
$ ,
which shows that Eq.~(18) is not satisfied. It is easily verified that
$$
X^\n\!_\aw = \d^\n_\aw - \d^\n_0 \d^2_\aw x^1   \tag29
$$
satisfies our condition
$
X^\m{}\!_\aw X^\aw{}\!_\n = \d^\m_\n
$.
If we multiply Eq.~(28) by Eq.~(29), we see that Eq.~(23) is satisfied.

\subhead
2.4. The Lagrangian
\endsubhead
We now recall (Pandres, 1999) evidence that the invariant $W_3 =
C^\n C_\n$ is an appropriate Lagrangian for gravitational and
electroweak unification.

The Riemann tensor is defined as usual by
$\dsize
R^\a{}\!_{\b\m\n}= h_i{}\!^\a \(h^i{}\!_{\b;\m;\n} - h^i{}\!_{\b;\n;\m}\)
$
while the Ricci tensor $R_{\m\n}$ and  Ricci scalar $R$
are defined, as usual, by $R_{\m\n}=R^\a{}\!_{\m\a\n}$ and $R=R^\a{}\!_\a$.
By using
$
h_i{}\!^\a h^i{}\!_{\b;\m;\n}=\(h_i{}\!^\a h^i{}\!_{\b;\m}\)_{;\n}
                              -h_i{}\!^\a{}\!_{;\n} h^i{}\!_{\b;\m}
                             = \g^\a{}\!_{\b\m;\n} +\g^\a{}\!_{\s\n} \g^\s{}\!_{\b\m}
$,
we easily find that
$$
R^\a{}\!_{\b\m\n}= \g^\a{}\!_{\b\m;\n} - \g^\a{}\!_{\b\n;\m}
                   +\g^\a{}\!_{\s\n} \g^\s{}\!_{\b\m}
                   -\g^\a{}\!_{\s\m} \g^\s{}\!_{\b\n} \quad .   \tag30
$$
From Eq.~(13), we see that
$
C_\m = h_i{}\!^\n\(h^i{}\!_\m,_\n-h^i{}\!_\n,_\m\)
     = h_i{}\!^\n\(h^i{}\!_{\m;\n}-h^i{}\!_{\n;\m}\)
     = \g^\n{}\!_{\m\n} - \g^\n{}\!_{\n\m}
     = \g^\n{}\!_{\m\n}
$ .
By using
$
C_\m = \g^\n{}\!_{\m\n}
$ ,
we find from Eq.~(30) that
$$
R_{\m\n}= C_{\m;\n}-C_\a \g^\a{}\!_{\m\n}- \g^\a{}\!_{\m\n;\a}
                                        +\g^\a{}\!_{\s\n} \g^\s{}\!_{\m\a} \quad ,     \tag31
$$
and, from Eq.~(31)
$$
C^\m C_\m = R + \g^{\m i\n} \g_{\m\n i}  - 2C^\m{}\!_{;\m} \quad . \tag32
$$
The first term on the right side of Eq.~(32) is the Ricci scalar, which
is the Lagrangian for gravitation. The last term is a covariant
divergence, which contributes nothing to the field equations. We now consider
the interpretation of the term $\g^{\m i\n} \g_{\m\n i}$.
From Eqs.~(7) and (8), we see that
$$
A^{\m\n i} M_{\m\n i}=0 \quad , \tag33
$$
and that
$$
M_{\m\n i}+M_{i \m\n}+M_{\n i\m}=0 \quad . \tag34
$$
From
$\dsize
\g_{\m\n i}=M_{\m\n i}+A_{\m\n i}
$ ,
and Eq.~(33), we get
$\dsize
\g^{\m i\n} \g_{\m\n i} =  M^{\m i\n} M_{\m\n i} - A^{\m\n i} A_{\m\n i}
$.
But,
$\dsize
M^{\m i\n} M_{\m\n i} =   {\tsize\frac{1}{2}} M^{\m i\n} M_{\m\n i}
                        + {\tsize\frac{1}{2}} M^{\n i\m} M_{\n\m i}
                      =   {\tsize\frac{1}{2}} M^{\m i\n} M_{\m\n i}
                        + {\tsize\frac{1}{2}} M^{i\n\m} M_{\m\n i}
                      =  {\tsize\frac{1}{2}} \(M^{\m i\n}
                        + M^{i\n\m}\) M_{\m\n i}
                      =  {\tsize\frac{1}{2}} M^{\m\n i} M_{\m\n i}
$ , where we have used Eq.~(34). Thus, we have
$\dsize
\g^{\m i\n} \g_{\m\n i} = {\tsize\frac{1}{2}} M^{\m\n i} M_{\m\n i} - A^{\m\n\a} A_{\m\n\a} .
$
We now define a vector
$$
A^\m = {\tsize\frac{1}{3!}}(-g)^{-1/2} e^{\m\a\b\s} A_{\a\b\s} \quad , \tag35
$$
and find that
$$
A^{\m\n\a} A_{\m\n\a}= -6 A^\m A_\m \quad .   \tag36
$$
In obtaining Eq.~(36), we have used the well known identity (see, e.g., Weber, 1961)
for expressing the product of two Levi-Civita symbols as a determinant of
Kronecker deltas. We now see that Eq.~(32) may be written
$$
C^\m C_\m = R +  {\tsize\frac{1}{2}} M^{\m\n i} M_{\m\n i} + 6 A^\m A_\m  - 2C^\m{}\!_{;\m} \quad . \tag37
$$
The term $M^{\m\n i} M_{\m\n i}$ is in the form of the usual electroweak  Lagrangian,
and the $A^\m A_\m$ term has precisely the form that is needed (see, e.g., Moriyasu,
1983) for the introduction of mass.

\subhead
2.5. Field Equations
\endsubhead
We have previously (Pandres, 1981) considered the variational principle
$\dsize
\d \I  C^\m C_\m \sq{-g}\, d^4\! x = 0
$
where $h^i{}\!_\m$ is varied.
We note that $\sq{-g}$ equals $h$, the determinant of $h^i{}\!_\m$; and that
$C^\m C_\m = C^i C_i$.
Hence, our variational principle may be written
$$
\d \I C^i C_i \, h \, d^4x = 0 \quad  .   \tag38
$$
The variational calculation (Pandres, 1984A) using $C^i C_i$ is less tedious than that
using $C^\m C_\m$.
We find from Eq.~(38) that
$$
\I h \(2 C^i \d C_i  - C^i C_i h^k{}\!_\n \d h_k{}\!^\n \)\, d^4x=0  \quad,  \tag39
$$
where we have used
$\d h = h h_k{}\!^\n \d h^k{}\!_\n = -h h^k{}\!_\n \d h_k{}\!^\n $.
We note that
$$\align
\(h h_i{}^\n\)\!,_\n =& h,_\n h_i{}^\n + h h_i{}^\n\!,_\n\\
       =& h \(h_k{}\!^\m h^k{}\!_{\m,\n} h_i{}^\n + h_k{}\!^\n{}\!,\!_\n h^k{}\!_\m h_i{}^\m\)\\
       =& h \(h_k{}\!^\n h^k{}\!_{\n,\m} h_i{}^\m - h_k{}\!^\n{} h^k{}\!_{\m,\n} h_i{}^\m\)\\
       =&-h C_\m h_i{}^\m = -h C_i\endalign
$$
Thus, we see that
$$
C_i =- h^{-1} \(h h_i{}\!^\n\)\!,_\n \quad  . \tag40
$$
Variation of Eq.~(40) gives
$\dsize
\d C_i = h^{-2} \( h h_i{}\!^\n \)\!,_\n  \d h
         -{h}^{-1} \d\(h h_i{}\!^\n\)\!,_\n
$\linebreak
$\dsize
= C_i h^k{}\!_\n \d h_k{}\!^\n -{h}^{-1} \[\d\(h h_i{}\!^\n\)\],_\n
$ .
Upon using this expression for $\d C_i$ in Eq.~(39), we obtain
$$
\I h C^k C_k  h^i{}\!_\n \d h_i{}\!^\n d^4x
- 2 \I C^i \[\d\(h h_i{}\!^\n\)\],_\n d^4x = 0  \quad  , \tag41
$$
and, integration by parts gives
$$
\I h \(C^i{}\!,_\n -h^i{}\!_\n C^k{}\!,_k
    +{\tsize\frac{1}{2}} h^i{}\!_\n C^k C_k\) \d h_i{}\!^\n d^4x
    -\I \[C^i \d\(h h_i{}\!^\n\)\],\!_\n d^4x=0 \quad  . \tag42
$$
By using Gauss's theorem, we may write the second integral of Eq.~(42) as an
integral over the boundary of the region of integration. We discard this boundary integral by demanding that
$C^i\d\(h h_i{}\!^\n\)$
shall vanish on the boundary,
and demand that $\d h_i{}\!^\n$ be arbitrary in the interior of the
(arbitrary) region of integration. Thus, we get field equations
$\dsize
C^i{}\!,_\n -h^i{}\!_\n C^k{}\!,_k
    +{\tsize\frac{1}{2}} h^i{}\!_\n C^k C_k=0
$ , and, upon multiplying by $h_j{}\!^\n$, we write these field equations as
$$
C^i{}\!,_j -\d^i_j C^k{}\!,_k
    +{\tsize\frac{1}{2}} \d^i_j C^k C_k=0  \quad . \tag43
$$
We note that
$
C^\a{}\!_{;\s}=\(C^k h_k{}\!^\a\)\!_{;\s}
               =C^k{}\!,_\s h_k{}\!^\a+ C^k h_k{}\!^\a{}\!_{;\s}
               =C^k{}\!,_\s h_k{}\!^\a+ C^k \g_k{}\!^\a{}\!_\s
$.
Thus, we have
$
C^k{}\!,_\s h_k{}\!^\a=C^\a{}\!_{;\s}+C^\r\g^\a{}\!_{\r\s}
$.
If we multiply by $h^i{}\!_\a h_j{}\!^\s$, we get
$
C^i{}\!,\!_j = h^i{}\!_\a h_j{}\!^\s \(C^\a{}\!_{;\s}+C^\r\g^\a{}\!_{\r\s}\)
$ ,
and
$
C^k{}\!,\!_k= C^\a{}\!_{;\a}+ C^\a C_\a
$ .
If we use these expressions for $C^i{}\!,\!_j$  and $C^k{}\!,\!_k$ in Eq.~(43), we obtain the relation
$
h^i{}\!_\a h_j{}\!^\s \(C^\a{}\!_{;\s}+C^\r\g^\a{}\!_{\r\s}\)
     -\d^i_j C^\a{}\!_{;\a}
    -{\tsize\frac{1}{2}} \d^i_j C^\a C_\a=0
$ ,
and, upon multiplying this by $h_{i\m} h^j{}\!_\n$, we rewrite our field equations as
$$
C_{\m;\n}-C_\a \g^\a{}\!_{\m\n}
     -g_{\m\n} C^\a{}\!_{;\a}
    -{\tsize\frac{1}{2}} g_{\m\n} C^\a C_\a=0 \quad . \tag44
$$
\subsubhead
2.5.1. The field equations as Einstein equations
\endsubsubhead
The Einstein equations of general relativity may be interpreted in two ways.
One interpretation is as differential equations for the metric, when the
stress-energy tensor is given. Alternatively, these equations may be looked
upon as a definition of the stress-energy tensor in terms of the metric.
The second interpretation has been stressed particularly by Schr\"odinger
(1960) [``I would rather you did not regard these equations as field equations, but
as a definition of $T_{ik}$ the matter tensor.''] and by Eddington (1924)
[``and we must proceed by inquiring first what experimental properties the
physical tensor possesses, and then seeking a geometrical tensor which
possesses these properties'']. It is the second interpretation that we adopt.

From Eqs.~(31) and (32), we find that an identity for the Einstein tensor
$
G_{\m\n} = R_{\m\n} - {\tsize\frac{1}{2}} g_{\m\n} R
$
is
$$\aligned
G_{\m\n} =& \,C_{\m;\n}-C_\a \g^\a{}\!_{\m\n}
     -g_{\m\n} C^\a{}\!_{;\a}
    -{\tsize\frac{1}{2}} g_{\m\n} C^\a C_\a \\
            & + \g_\m{}\!^\a{}\!_{\n;\a} + \g^\a{}\!_{\s\n} \g^\s{}\!_{\m\a}
              + {\tsize\frac{1}{2}} g_{\m\n} \g^{\a i\s} \g_{\a\s i} \quad .
\endaligned \tag45
$$
Equation (44) just states that the first line on the right side of Eq.~(45) vanishes.
Thus, we may write our field equations as
$$
G_{\m\n} =  \g_\m{}\!^\a{}\!_{\n;\a} + \g^\a{}\!_{\s\n} \g^\s{}\!_{\m\a}
              + {\tsize\frac{1}{2}} g_{\m\n} \g^{\a i\s} \g_{\a\s i} \quad . \tag46
$$
By using the well known symmetry of the Einstein tensor, i.e., $G_{\m\n}=G_{\n\m}$.
we see from Eq.~(46) that the symmetric part of our field equations is
$$
G_{\m\n} = {\tsize\frac{1}{2}} \(\g_\m{}\!^\a{}\!_\n  +  \g_\n{}\!^\a{}\!_\m\)\!_{;\a}
 + {\tsize\frac{1}{2}}\(\g^\a{}\!_{\s\n} \g^\s{}\!_{\m\a}+ \g^\a{}\!_{\s\m} \g^\s{}\!_{\n\a}\)
             + {\tsize\frac{1}{2}} g_{\m\n} \g^{\a i\s} \g_{\a\s i} \quad . \tag47
$$
Since
$
\g_\m{}\!^\a{}\!_\n =  M_\m{}\!^\a{}\!_\n + A_\m{}\!^\a{}\!_\n
$ ,
we see that
$
\(\g_\m{}\!^\a{}\!_\n + \g_\n{}\!^\a{}\!_\m\)_{;)\a}
= \(M_\m{}\!^\a{}\!_\n + M_\n{}\!^\a{}\!_\m\)_{;\a}
= \(M_\m{}\!^\a{}\!_i h^i{}\!_\n  + M_\n{}\!^\a{}\!_i h^i{}\!_\m\)_{;\a}
$
$
= J_{\m i} h^i{}\!_\n + J_{\n i} h^i{}\!_\m
  + M_\m{}\!^\a{}\!_\s \g^\s{}\!_{\n\a} + M_\n{}\!^\a{}\!_\s \g^\s{}\!_{\m\a}
$ ,
where\linebreak
\flushpar
$
J_{\m i} = M_\m{}\!^\a{}\!_{i;\a}
$
is a (conserved) electroweak current. From Eq.~(33), the repeated use of Eq.~(34),
the total antisymmetry of $A_{\m\n\a}$, and the antisymmetries of $\g_{\m\n\a}$ and
$M_{\m\n\a}$ in their first two indices, we find after a tedious but straightforward
calculation that Eq.~(47) may be written
$$
G_{\m\n}= A^{ij}{}\!_\m A_{ij\n} - {\tsize\frac{1}{2}} g_{\m\n} A^{ij\a} A_{ij\a}
         + {\tsize\frac{1}{2}}\( J_{\m i} h^i{}\!_\n + J_{\n i} h^i{}\!_\m\) - M_{\m\n}
\quad , \tag48
$$
where $\dsize M_{\m\n} = M^\a{}\!_{\m i} M_{\a\n}{}\!^i - {\tsize\frac{1}{4}} g_{\m\n}
M^{\a\s i} M_{\a\s i} $ . The terms in Eq.~(48) that involve $A_{ij\m}$ may be written
in a more simple form. From Eq.~(35), we have $\dsize A_\m =
{\tsize\frac{1}{3!}}(-g)^{-1/2} g_{\m\r} e^{\r\a\b\s} A_{\a\b\s} $ , and we find that
$\dsize A_\m = -{\tsize\frac{1}{3!}}(-g)^{1/2} e_{\m\a\b\s} A^{\a\b\s} $. Thus, $\dsize
A_\m A_\n = -{\tsize\frac{1}{36}} g_{\m\r} e^{\r\a\b\s} e_{\n\t\l\y} A^{\t\l\y}
A_{a\b\s} $. By expressing the product of Levi-Civita symbols as a determinant of
Kronecker deltas, we get $\dsize A_\m A_\n = $ \newline $\dsize {\tsize\frac{1}{2}}
A^{ij}{}\!_\m A_{ij\n}
            - {\tsize\frac{1}{6}} g_{\m\n} A^{ij\a} A_{ij\a}
$.
From this and Eq.~(36), we see that Eq.~(48) may be written
$$
G_{\m\n}= 2A_\m A_\n + g_{\m\n} A^\a A_\a
         + {\tsize\frac{1}{2}}\( J_{\m i} h^i{}\!_\n + J_{\n i} h^i{}\!_\m\) - M_{\m\n}
\quad . \tag49
$$
The right side of Eq.~(49) is just what one would expect for the stress-energy tensor of the
electroweak field, its associated currents, and gauge symmetry breaking terms corresponding
to those in the Lagrangian, Eq.~(37).

\subhead
2.6. Solutions of the field equations
\endsubhead
\subsubhead
2.6.1. Solutions with  $C_i=0$
\endsubsubhead
It is clear that our field equations, Eq.~(43) are satisfied if $C_i=0$. Consider the tetrad $h^i{}\!_\m = \d^i{}\!_\m + \d^i{}\!_0 \d^2{}\!_\m x^1$, where $x^1$ is a Greek (space-time) coordinate. We have shown (Pandres, 1981) that this tetrad yields $C_i=0$, gives a Ricci scalar
$\dsize
R={\tsize\frac{1}{2}}
$, and gives a metric $g_{\m\n}$ which satisfies the well known (Synge, 1960) Einstein equations for a charged dust cloud.

\subsubhead
2.6.2.
Solutions with $C_i$ constant and lightlike
\endsubsubhead
It is also clear that our field equations are satisfied if $C_i$ is constant and lightlike. Consider the tetrad
$$
h^i_{}\!_\m = \d^i_\m + \(\d^i_0 + \d^i_1\)\d^0_\m \(e^{x^1} -1\)  \tag50
$$
where the coordinate $x^1$ is Greek.
We have shown (Pandres, 1984A) that this tetrad yields a nonvanishing
but constant and lightlike $C_i$.

\subsubhead
2.6.3.
Solutions with $C_i$ which does not vanish, and is neither constant nor lightlike.
\endsubsubhead
It is clear from Secs.~2.6.1 and ~2.6.2 above that a tetrad satisfies our field equations if it satisfies the condition of either vanishing or being constant and lightlike. In a previous paper, (Pandres, 1984A), we made the false assertion that a tetrad satisfies our field equations {\sl only} if it satisfies this condition. The false assertion was based on the following argument: It is clear that for distinct values of $i$ and $j$, the field equations state that $C^i{}\!,_j=0$. This fact led us to assume that the component $C^i$ can depend only on the single coordinate $x^i$; i.e., that $C^0{}\!,_0$ can depend only on $x^0$; $C^1{}\!,_1$ only on $x^1$, etc. This assumption would be true if the Latin coordinates were holonomic, but is false, because they are nonholonomic. An example of a tetrad which satisfies our field equations, but yields a $C_i$ which is neither constant nor lightlike has been found by one of us (Green).  His tetrad is
$$
h_i{}\!^\m = \[\(x^0\)^2 - \(x^1\)^2 + \(x^2\)^2 - \(x^3\)\]\d^\m_i \; .   \tag51
$$
In Eq.~(51), the coordinates \;
$\dsize
x^0,x^1,x^2,x^3
$
\; are Greek. The tetrad in Eq.~(51) satisfies our field equations, but yields
$\dsize
C_i = \(-6x^0,-6x^1,-6x^2,-6x^3\)
$ which is neither constant nor lightlike.

\subsubhead
2.6.4. Solutions that yield flat Riemann space-times
\endsubsubhead
We note that our field equations admit non-trivial solutions for which $g_{\m\n}$ is the metric of a flat space-time.
One of us (Green, 1991) has exhibited the tetrad
$$
h_i{}^\m
=\d^\m_0 \d^0_i+\d^\m_3 \d^3_i
+\(\d^\m_1 \d^1_i+\d^\m_2 \d^2_i\)\cos x^3
+\(\d^\m_2 \d^1_i-\d^\m_1 \d^2_i\)\sin x^3 \;  ,\tag52
$$
where the coordinate $x^3$ is Greek.
For this $h_i{}^\m$, the quantity $M_{\m\n i}$ does not vanish, but $C_i = 0$, and
$g_{\m\n} = \text{diag}(-1,1,1,1)$.
He has also exhibited (Green, 1997) the tetrad
$$\aligned
h_i{}^\m = &{\tsize\frac{1}{2}}
             \[\(\d_0^\m  \d_i^0  + \d_1^\m  \d_i^1 \)\(F + \frac{1}{F}\)
            + \(\d_0^\m  \d_i^1  + \d_1^\m  \d_i^0 \)\(F - \frac{1}{F}\)\]\\
            &+  \d_2^\m  \d_i^2  + \d_3^\m  \d_i^3  ,
\endaligned\tag53$$
where $F=x^0+x^1$, and the coordinates $x^0$ and $x^1$ are Greek. For this $h_i{}^\m$,
the quantity $M_{\m\n i}$ does not vanish, but $C_i$ is constant and lightlike, and
$g_{\m\n} = \text{diag}(-1,1,1,1)$.

\subhead
3. THE HAMILTONIAN
\endsubhead
In this section the Hamiltonian for this
theory will be derived.  As the dynamics is constrained, the Dirac-Bergmann
procedure will be used to find all the constraints and to produce a
consistent Hamiltonian.

\subsubhead
3.1. The Primary Hamiltonian
\endsubsubhead
Let $h$ be the determinant of $h^i_{\text{ }\mu}$
as before.  The 16 canonical position variables are defined by
$$Q_i^{\alpha}\equiv h\text{ }h_i^{\text{ }\alpha} . \tag{54}$$
Let $Q^i_\alpha$ be the matrix
inverse of $Q_i^\alpha$, so that $Q_i^\alpha Q^i_\beta = \delta_\beta^\alpha$
and $Q^i_\alpha Q_j^\alpha=\delta^i_j$.
Let $Q$ be the determinant of $Q_i^\alpha$.  Then
$Q=\text{det}(Q_i^\alpha)=\text{det}(h\text{ } h_i^{\text{ }\alpha}) = h^4\cdot h^{-1} =
h^3 \,$, and so, $h= Q^{\frac13}$.
Using Eq.~(40) we have
$$C_i = -Q^{-\frac13}Q_{i,\nu}^\nu \quad ,\tag{55}$$
and therefore the Lagrangian density may be expressed by
$$\text{\it{L}}=g^{ij}Q_{i,\mu}^\mu Q_{j,\nu}^\nu \,Q^{-\frac13}
\quad . \tag{56}$$
The momenta are defined as usual by
$P^i_\mu=\dfrac{\delta L}{\delta Q_{i,0}^\mu}$ where $Q_{i,0}^\mu$ is the
derivative of $Q_i^\mu$ with respect to the Greek $x^0$ variable.  We assume
that the $x^0$ variable has a time-like direction at each point in space-time.
We also assume that the values of $Q_i^\alpha$ and $P^i_\mu$ as well as their
derivatives on a space-like surface $\sigma$ determine the dynamics. From
Eq.~(56) we have
$$P^i_\mu=2g^{ij} Q_{j,\nu}^\nu \delta^0_\mu \text{ }Q^{-\frac13}\, . \tag{57}$$
For the remainder of this section we will use a bar over an index to indicate
a restriction of the index range to the values 1, 2, and 3.  Thus there are 12
primary constraints
$$P^i_{\bar{\mu}}= 0 \qquad \bar{\mu}=1,2,3 \quad . \tag{58}$$
The 4 nonzero momenta are seen to be a multiple of the Latin components
of the curvature vector:
$$P^i_0=2g^{ij}Q_{j,\nu}^\nu \,Q^{-\frac13}=
-2g^{ij}C_j=-2C^i \quad . \tag{59}$$
The Hamiltonian density {\it H} is defined by $\text{\it H} =
P^i_\mu Q_{i,0}^\mu-\text{\it L}$. Using the constraints and Eq.~(59), we have
$$\text{\it H}=P^i_0 Q_{i,0}^0 - \frac14g_{ij}P^i_0 P^j_0 Q^{\frac13}
$$
But using Eq.~(55), $Q_{i,0}^0=
Q_{i,\mu}^\mu-Q_{i,\bar{\mu}}^{\bar{\mu}}=
-Q^{\frac13}C_i - Q_{i,\bar{\mu}}^{\bar{\mu}} =
\frac12Q^{\frac13}g_{ij}P^j_0 - Q_{i,\bar{\mu}}^{\bar{\mu}}$, and therefore
$$\aligned \text{\it H}=&
\frac12g_{ij}P^i_0P^j_0Q^{\frac13} -
P^i_0 Q_{i,\bar{\mu}}^{\bar{\mu}} -
\frac14 g_{ij}P^i_0 P^j_0 Q^{\frac13}  \\
=&\frac14g_{ij}P^i_0P^j_0Q^{\frac13} -
P^i_0 Q_{i,\bar{\mu}}^{\bar{\mu}} \qquad .  \endaligned  $$
Hence we have the following primary Hamiltonian density
$$H_p=\frac14g_{ij}P^i_0 P^j_0 Q^{\frac13} - P^i_0 Q_{i,\bar{\mu}}^{\bar{\mu}}
+v_i^{\bar{\mu}} P^i_{\bar{\mu}} \tag{60}$$ with Lagrange
multipliers $v_i^{\bar{\mu}}$. In cases where $C^i$ is zero on the
boundary of $\sigma$, a partial integration yields the following
primary Hamiltonian density:
$$\text{\it H}_p =
\frac14g_{ij}P^i_0P^j_0Q^{\frac13}
+ Q_i^{\bar{\mu}} P^i_{0,\bar{\mu}}
+ v^{\bar{\mu}}_i P^i_{\bar{\mu}} \qquad . \tag{61} $$
The full Hamiltonian is $\Cal H_p = {\dsize \int_{\sigma} H_p \, d^3 x}$.

\subsubhead
3.2. The Dirac-Bergmann Procedure and the Consistent Hamiltonian
\endsubsubhead \flushpar
When the dynamics are constrained,
consistency requires that the time derivative of a constraint must be zero.
We follow the Dirac-Bergmann algorithm (Bergmann and Goldberg, 1955),
(Dirac, 1964), (Sundermeyer, 1982), (Weinberg, 1995)
for constrained dynamics.
The Hamiltonian equations of motion state that for any function $X$ of
$Q_i^\mu$ and $P^i_\mu$ we have
$$\dfrac{dX}{dt}=\{X,H_p\}=
\frac{\delta X}{\delta Q_i^\mu}\frac{\delta H_p}{\delta P^i_\mu}-
\frac{\delta X}{\delta P^i_\mu}\frac{\delta H_p}{\delta Q_i^\mu}
\quad  . $$ Dirac refers to the constraint equations as weak
equations, since one must be careful to use these equations only
after time derivatives and other variations are calculated.  We
shall use $x$ to represent the variable of integration for the
first functional in the bracket and $y$ to represent the variable
of integration for the second functional.  When there is no
possible confusion these variables will be suppressed.  We will
also suppress display of the integrals over $\sigma$ and only show
the integrands.

Since, in general, $C_i$ may have any value on the boundary, we
proceed from equation Eq.~(60). Substituting $W =
g_{ij}P^i_0P^j_0$, and noting that $\dfrac{\delta Q}{\delta
Q_i^\beta}=QQ^i_\beta$, we have
$$\frac{dP^i_{\bar{\alpha}}}{dt}=-\frac{\delta H_p}{\delta Q_i^{\bar{\alpha}}}
=-\frac1{12}WQ^{\frac13}Q^i_{\bar{\alpha}}\delta(x-y)+P^i_0\delta_{,\bar{\alpha}}(x-y)
$$
and thus for consistency
$$\frac1{12}WQ^{\frac13}Q^i_{\bar{\alpha}}\delta(x-y)-
P^i_0\delta_{,\bar{\alpha}}(x-y)=0 \qquad ,$$
which becomes after integration with respect to the $y$ variable
$$\frac1{12}WQ^{\frac13}Q^i_{\bar{\alpha}}+P^i_{0,\bar{\alpha}}=0 \qquad .
\tag{62}$$
We assume fixed boundary conditions so that the variation on the boundary
is zero and thus the boundary term is zero.
Equation Eq.~(62) represents 12 secondary constraints on the theory.

Before proceeding with the constraint algorithm a comment is in order.
Computation of $P^i_{0,0}=[P^i_0,H_p]$, yields
$P^i_{0,0}=-\frac1{12}WQ^\frac13Q^i_0$.  Hence, we have
$\frac1{12}WQ^\frac13Q^i_\alpha+P^i_{0,\alpha}=0$ for $\alpha=0, 1, 2, 3$.
Since $Q^\frac13Q^i_\alpha=h^i_\alpha$ and $P^i_0=-2C^i$ and $W=4C^kC_k$
we see that Eq.~(62) along with the dynamics for $P^i_0$ imply that
$C^i_{,j}=\frac16\delta^i_jC^kC_k$ which is the Latin form of the field
equations.

Proceeding with the algorithm we note that multiplication of Eq.~(62) by
$Q_i^0(x)$ yields the condition
$$Q_i^0P^i_{0,\bar{\alpha}} = 0 \qquad . \tag{63}$$
The computation of further constraints is rather tedious.  It will be useful
to note that
$\dfrac{\delta Q^i_\alpha}{\delta Q^\beta_j}=-Q^i_\beta Q^j_\alpha$.  Because of
Eq.~(63) we must require that
$$\aligned  0=&[Q_i^0P^i_{0,\bar{\alpha}},H_p]\\
=&P^i_{0,\bar{\alpha}} \Biggl(\frac12g_{ij}P^j_0Q^\frac13-Q^{\bar{\gamma}}_{i,\bar{\gamma}}
\Biggr)\delta - Q_i^0(x)\delta_{,\bar{\alpha}}(x-y)
\cdot\Biggl(\frac12g_{jk}P^j_0P^k_0Q^\frac13Q^i_0\Biggr)(y)
\endaligned$$
After the integration with respect to $y$ we have
$$\aligned
0=&\frac12g_{ij}P^i_{0,\bar{\alpha}}P^j_0Q^\frac13
-P^i_{0,\bar{\alpha}}Q^{\bar{\gamma}}_{i,\bar{\gamma}}+
\frac1{12}\biggl(g_{jk}P^j_0P^k_0Q^\frac13Q^i_0\biggr)_{,\bar{\alpha}}Q_i^0\\
=&\frac23g_{ij}P^i_{0,\bar{\alpha}}P^j_0Q^\frac13
-P^i_{0,\bar{\alpha}}Q^{\bar{\gamma}}_{i,\bar{\gamma}}
+\frac1{36}WQ^\frac13Q^k_\beta Q^\beta_{k,\bar{\alpha}}
+\frac1{12}WQ^\frac13Q^i_{0,\bar{\alpha}}Q_i^0 \\
=&\frac1{36}WQ^\frac13\bigl(-2g_{ij}Q^i_{\bar{\alpha}}P^j_0Q^\frac13
+3Q^i_{\bar{\alpha}}Q^{\bar{\gamma}}_{i,\bar{\gamma}}
+Q^i_\beta Q^\beta_{i,\bar{\alpha}}
-3Q^i_0Q^0_{i,\bar{\alpha}} \bigr) \quad .
\endaligned$$
In the last line, the secondary  constraint Eq.~(62) has been used.  Since $Q=0$
is not acceptable, we find that
$$W=0 \qquad \text{  or  } \qquad -2g_{ij}Q^i_{\bar{\alpha}}P^j_0Q^\frac13
+3Q^i_{\bar{\alpha}}Q^{\bar{\gamma}}_{i,\bar{\gamma}}
+Q^i_\beta Q^\beta_{i,\bar{\alpha}}
-3Q^i_0Q^0_{i,\bar{\alpha}}=0 \quad .\tag{64a,b}$$

{\bf Type I Regions.}  We define Type I regions as simply connected regions
of the space-like surface $\sigma$ where $W\equiv 0$.  In Type I regions the
secondary constraint Eq.~(62) implies that $P^i_{0,\bar{\alpha}}=0$ also.  We
then require that
$$0=[W,H_p]=-2g_{ij}P^j_0\frac{\delta H_p}{\delta Q_i^0}=-2g_{ij}P^j_0\cdot
\frac1{12}WQ^\frac13 Q^i_0$$
and since $W$ is assumed to be zero, this is automatically satisfied.  Thus
in the $W=0$ case, the algorithm terminates.  The secondary constraints in
this case may be summarized by the 4 conditions:
$$P^i_0=K^i \quad \text{, where }K^i \text{ is constant and lightlike.} \tag{65}$$
These represent 4 first class, secondary constraints.

{\bf Type II Regions.}  We define Type II regions as simply connected regions
of $\sigma$ where $W$ is not identically zero.   In this case we assume that
$-2g_{ij}Q^i_{\bar{\alpha}}P^j_0Q^\frac13
+3Q^i_{\bar{\alpha}}Q^{\bar{\gamma}}_{i,\bar{\gamma}}
+Q^i_\beta Q^\beta_{i,\bar{\alpha}}
-3Q^i_0Q^0_{i,\bar{\alpha}} $ is identically zero.  Returning to the constraint
given by Eq.~(62) we require that
$$\aligned 0=
&\biggl[\frac1{12}WQ^\frac13Q^i_{\bar{\alpha}}+P^i_{0,\bar{\alpha}}\, ,\,H_p\biggr] \\
=&-\frac1{36}WQ^\frac13\bigr(2Q^i_0g_{kl}Q^k_{\bar{\alpha}}P^l_0Q^\frac13
-3Q^i_0Q^k_{\bar{\alpha}}Q^{\bar{\gamma}}_{k,\bar{\gamma}}
+3Q^i_{\bar{\gamma}}Q^k_{\bar{\alpha}}v^{\bar{\gamma}}_k\\
& \qquad \qquad \quad +Q^i_{\bar{\alpha}}Q^k_0Q^{\bar{\gamma}}_{k,\bar{\gamma}}
-Q^i_{\bar{\alpha}}Q^k_{\bar{\gamma}}v^{\bar{\gamma}}_k
-Q^k_\gamma Q^\gamma_{k,\bar{\alpha}}Q^i_0
-3Q^i_{0,\bar{\alpha}} \bigr)
\endaligned $$
If we assume that W and Q are not identically zero, then we have 12 tertiary
constraints:
$$ \aligned
0=&2Q^i_0g_{kl}Q^k_{\bar{\alpha}}P^l_0Q^\frac13
-3Q^i_0Q^k_{\bar{\alpha}}Q^{\bar{\gamma}}_{k,\bar{\gamma}}
+3Q^i_{\bar{\gamma}}Q^k_{\bar{\alpha}}v^{\bar{\gamma}}_k
+Q^i_{\bar{\alpha}}Q^k_0Q^{\bar{\gamma}}_{k,\bar{\gamma}} \\
& \qquad -Q^i_{\bar{\alpha}}Q^k_{\bar{\gamma}}v^{\bar{\gamma}}_k
-Q^k_\gamma Q^\gamma_{k,\bar{\alpha}}Q^i_0
-3Q^i_{0,\bar{\alpha}} \endaligned \tag{66}$$
By multiplication by appropriate factors of $Q_i^\beta$ we may split these 12
equations as follows.  Multiplication of Eq.~(66) by $Q_i^0$ implies that
$2g_{ij}Q^i_{\bar{\alpha}}P^j_0Q^\frac13
-3Q^k_{\bar{\alpha}}Q^{\bar{\gamma}}_{k,\bar{\gamma}}
-Q^k_\beta Q^\beta_{k,\bar{\alpha}}
+3Q^k_0Q^0_{k,\bar{\alpha}}=0$.  This is equivalent to the 3 constraints given
in Eq.~(64b).  Next, multiplication of Eq.~(66) by $Q_i^{\bar{\alpha}}$ yields the
single constraint
$$Q^k_0Q^{\bar{\gamma}}_{k,\bar{\gamma}}=0 \quad . \tag{67}$$
Finally, multiplication of Eq.~(66) by $Q_i^{\bar{\beta}}$ results in the
conditions  $Q^k_{\bar{\alpha}}v^{\bar{\beta}}_k
-\frac13\delta^{\bar{\beta}}_{\bar{\alpha}}Q^k_{\bar{\gamma}}v^{\bar{\gamma}}_k
+Q^i_0Q^{\bar{\beta}}_{i,\bar{\alpha}} = 0$.  Using Eq.~(67) these 9 equations are
seen to be traceless and hence these 8 equations may be used to reduce the
number of unknown Lagrange multipliers from 12 to 4.  The result is
$$v_k^{\bar{\beta}}=\lambda Q_k^{\bar{\beta}}+\lambda^{\bar{\beta}}Q_k^0
-Q^j_0Q^{\bar{\beta}}_{j,\bar{\gamma}}Q^{\bar{\gamma}}_k \qquad , \tag{68}$$
where $\lambda$ and $\lambda^{\bar{\beta}}$ represent 4 arbitrary Lagrange
multiplier functions.

It follows from Eq.~(67) that we must require
$$\aligned
0=&\biggl[Q^k_0Q^{\bar{\gamma}}_{k,\bar{\gamma}}\, , \, H_p\biggr]\\
=&Q^{\bar{\gamma}}_{k,\bar{\gamma}}\frac{\delta Q^k_0}{\delta Q_l^\alpha}
\Biggl(\delta_0^\alpha\biggl(\frac12g_{il}P^i_0Q^\frac13-
Q^{\bar{\beta}}_{l,\bar{\beta}}\biggr)
+v_l^{\bar{\alpha}}\Biggr)+Q^k_0\frac{\delta Q^{\bar{\gamma}}_{k,\bar{\gamma}} }
{\delta Q_l^{\bar{\alpha}} }v_l^{\bar{\alpha}}\\
=&Q^{\bar{\gamma}}_{k,\bar{\gamma}}\Biggl(-Q^k_\alpha Q^l_0\Biggr)\Biggl(
\delta_0^\alpha\biggl(\frac12g_{il}P^i_0Q^\frac13-Q^{\bar{\beta}}_{l,\bar{\beta}}
\biggr)+v_l^{\bar{\alpha}}\Biggr)\\ & \qquad +Q_0^k(x)\delta^{\bar{\gamma}}_{\bar{\alpha}}
\delta_k^l \delta_{,\bar{\gamma}}(x-y)v_l^{\bar{\alpha}}(y)\\
=&-Q^{\bar{\gamma}}_{k,\bar{\gamma}}Q^k_{\bar{\alpha}}Q^l_0v_l^{\bar{\alpha}}
-Q^l_0v^{\bar{\alpha}}_{l,\bar{\alpha}} \qquad ,
\endaligned$$
where the constraints have been used and an integration by
parts has been performed on the second term.  Now using Eq.~(68) we find
$$\lambda^{\bar{\beta}}_{,\bar{\beta}}=
-\bigl(Q^k_{\bar{\beta}}Q^{\bar{\gamma}}_{k,\bar{\gamma}}+Q^k_0Q^0_{k,\bar{\beta}}
\bigr)\lambda^{\bar{\beta}}
+Q^k_0Q^{\bar{\beta}}_{k,\bar{\gamma}}Q^l_0Q^{\bar{\gamma}}_{l,\bar{\beta}}
\tag{69}$$
This differential equation represents one condition on the 3 multipliers
$\lambda^{\bar{\beta}}$.

Finally we proceed from the constraint given in Eq.~(64b).
$$\aligned
0=&\biggl[-2g_{ij}Q^i_{\bar{\alpha}}P^j_0Q^\frac13
+3Q^i_{\bar{\alpha}}Q^{\bar{\gamma}}_{i,\bar{\gamma}}
+Q^k_\beta Q^\beta_{k,\bar{\alpha}}
-3Q^i_0Q^0_{i,\bar{\alpha}} \, , \, H_p \biggr]\\
=&\Biggl(-2g_{kl}P^l_0Q^k_\beta Q^j_{\bar{\alpha}}Q^\frac13+
\frac23g_{kl}P^l_0Q^k_{\bar{\alpha}}Q^j_\beta Q^\frac13
+3Q^{\bar{\gamma}}_{k,\bar{\gamma}}Q^k_\beta Q^j_{\bar{\alpha}}
-3Q^j_{\bar{\alpha}}\delta_{,\bar{\beta}}\\
& \qquad +Q^\gamma_{k,\bar{\alpha}}Q^k_\beta Q^j_\gamma
-Q^j_\beta \delta_{,\bar{\alpha}}
-3Q^0_{i,\bar{\alpha}}Q^i_\beta Q^j_0
+3Q^j_0 \delta_{,\bar{\alpha}} \delta^0_\beta \Biggr) \\
& \qquad \cdot \Biggl(\delta^\beta_0\biggl(\frac12g_{jm}P^m_0Q^\frac13
-Q^{\bar{\gamma}}_{j,\bar{\gamma}}\biggr) + v_j^{\bar{\beta}}\Biggr)\\
&\qquad - \frac16g_{kl}Q^k_{\bar{\alpha}}Q^l_0g_{mn}P^m_0P^n_0Q^\frac23
\endaligned$$
Substituting for $v_j^{\bar{\beta}}$ using Eq.~(68) and using the constraints Eqs.~(62), ~(64b) and ~(67) yields
$$\aligned
0=&g_{kl}Q^k_0P^l_0Q^\frac13\biggl(Q^i_{\bar{\alpha}}Q^{\bar{\beta}}_{i,\bar{\beta}}
-\frac23Q^i_\beta Q^\beta_{i,\bar{\alpha}}-\frac12Q^i_0Q^0_{i,\bar{\alpha}}\biggr)
-\frac1{12}g_{kl}P^k_0P^l_0g_{ij}Q^i_{\bar{\alpha}}Q^j_0Q^\frac23 \\
&\qquad +5Q^i_{0,\bar{\alpha}}Q^{\bar{\beta}}_{i,\bar{\beta}}
-3Q^i_{0,\bar{\beta}}Q^{\bar{\beta}}_{i,\bar{\alpha}}
+2g_{kl}Q^k_{\bar{\beta}}P^l_0Q^i_0Q^{\bar{\beta}}_{i,\bar{\alpha}}Q^\frac13 \\
& \qquad -Q^i_0Q^{\bar{\beta}}_{i,\bar{\gamma}}\biggl(3Q^j_{\bar{\alpha}}Q^{\bar{\gamma}}_{j,\bar{\beta}}
+2Q^j_{\bar{\beta}}Q^{\bar{\gamma}}_{j,\bar{\alpha}}\biggr)
-\frac12g_{kl}P^k_0Q^l_{0,\bar{\alpha}}Q^\frac13 \\
& \qquad +\biggl(6Q^i_{\bar{\alpha}}Q^{\bar{\beta}}_{i,\bar{\beta}}
  +2Q^i_{\bar{\beta}}Q^{\bar{\beta}}_{i,\bar{\alpha}} \biggr)\lambda
+\biggl(3Q^i_{\bar{\alpha}}Q^0_{i,\bar{\beta}}
-Q^i_{\bar{\beta}}Q^0_{i,\bar{\alpha}} \biggr)\lambda^{\bar{\beta}} \\
& \qquad +6\lambda_{,\bar{\alpha}}
\endaligned \tag{70} $$
These 3 equations along with Eq.~(69) may be used to solve for
$\lambda$ and the $\lambda^{\bar{\beta}}$ and since these are
first order differential equations in the lambdas, we expect that
there will be 4 arbitrary constants in our solutions for the
Lagrange multipliers. This completes the Dirac-Bergmann algorithm.
For a summary see Table I.

For tetrads that satisfy the field equations we may check to determine whether
the tetrad also agrees with the results of the Dirac-Bergmann algorithm.
For tetrads with
$C^i=0$ or $C^i$ constant and lightlike these results are clearly
consistent and the region is Type I.  When $C^i$ is nonconstant and the field
equations are satisfied
it is not so obvious because all the tertiary constraints must be checked.
For the example given in Eq.~(51), one finds that
the tertiary constraints are indeed satisfied and the solutions for the Lagrange
multipliers are $\lambda=-\dfrac{16x^0}{7\phi}+\dfrac{\kappa^0x^0}{\phi^6}$
and $\lambda^{\bar{\alpha}}=\kappa^{\bar{\alpha}}\phi^6$, where
$\phi=(x^0)^2-(x^1)^2-(x^2)^2-(x^3)^2$, and $\kappa^\alpha$ are 4 arbitrary
constants.

At present we do not have a physical interpretation of all the constraints.
Recall that Gauss's law shows up as one of the constraints in the free
electromagnetic field (Dirac, 1964).  We expect that our secondary
constraints will also have a similarly important interpretations.

In the case of Type I regions with $P^i_0=0$ we see that the Hamiltonian is
consistent with what is expected in a theory that describes gravitation.
Multiplication of Eq.~(62) by $Q^{\bar{\alpha}}_i$ implies that
$\frac14WQ^\frac13+Q_i^{\bar{\alpha}}P^i_{0,\bar{\alpha}}=0$.  By comparison
to Eq.~(61) we see that $H_p$ is weakly zero if $P^i_0=-2C^i$ is zero (Type I).
Many investigators (e.g., Misner, 1957) expect that the correct Hamiltonian
for gravity should be weakly zero.

Of the two types of regions, it would seem that the Type I regions would be
more physically relevant.  The 16 first class constraints would correspond
to 16 gauge degrees of freedom.  All the constraints are first class
so that there is no need for the Dirac bracket.
The Type II regions, however, have no gauge degrees of freedom and
the Dirac bracket would be needed to define the symplectic structure on the
4-dimensional phase space.

\vskip 0.5in

\heading Table I \endheading

$$\matrix
\text{CASE:} & W=0 & W\neq 0\\
 & & \\
\text{Primary} & P^i_{\bar{\alpha}}=0 & P^i_{\bar{\alpha}}=0 \\
\text{Constraints:}&\text{(12 First Class)} &\text{(12 Second Class)} \\
 & & \\
\text{Secondary} & P^i_0=K^i\text{, with}  &
\frac1{12}WQ^\frac13 Q^i_{\bar{\alpha}}+P^i_{0,\bar{\alpha}}=0 \\
\text{Constraints:}&\text{constant }K^i\text{ lightlike} & \text{(12 Second Class)} \\
&\text{(4 First Class)} & \\
 & & \\
\text{Tertiary} & \text{None} & Q^k_0Q^{\bar{\beta}}_{k,\bar{\beta}}=0 \text{ and}\\
\text{Constraints:} & & 2g_{ij}Q^i_{\bar{\alpha}}P^j_0Q^\frac13-3Q^k_{\bar{\alpha}}Q^{\bar{\beta}}_{k,\bar{\beta}} \\
 & & \qquad -Q^k_\beta Q^\beta_{k,\bar{\alpha}}+3Q^k_0Q^0_{k,\bar{\alpha}} = 0 \\
  & & \text{ (4 Second Class)} \\
  & & \\
\text{Gauge Fixing} & \text{16 required} & \text{Gauge fixed by} \\
\text{Constraints:} & & \text{constraint algorithm} \\
 & & \\
\text{Degrees of} & & \\
\text{Freedom} & 0 & 4 \\
\endmatrix
$$

\vskip 0.7in

\subhead
4. CONCLUDING REMARKS
 \endsubhead
\subsubhead
4.1 Possible Inclusion of the Strong Interaction
\endsubsubhead

It may be possible to extend our theory to include the strong interaction, by replacing the real orthonormal tetrad $h^i{}\!_\m$ with a complex orthonormal tetrad $Z^i{}\!_\m$ which is restricted so that the space-time metric
$$
g_{\m\n} = g_{ij} \overline {Z^i{}\!_\m} Z^j{}\!_\n    \quad \tag71
$$
remains {\sl real}. A bar indicates complex conjugation. That there exist complex tetrads which yield real metrics may be seen in the following way. It is known (see, e.g., Barut, 1980) that there exist {\sl two} complex groups which preserve the canonical Lorentz metric. One of these groups has complex transformation coefficients $t^i{}\!_m$ which satisfy the relation
$\dsize
g_{mn} = g_{ij} t^i{}\!_m t^j{}\!_n
$
where
$g_{ij}= g_{mn}=\text{diag}(-1,1,1,1)$. This group does not
contain $SU(3)$ as a subgroup, and hence is of no interest here.
The other group has complex transformation coefficients $T^i{}\!_m$ which satisfy the relation
$$
g_{mn} = g_{ij} \overline {T^i{}\!_m} T^j{}\!_n \tag72
$$
where a bar indicates complex conjugation. This group contains
$SU(3)$ as a proper subgroup. The components of the complex tetrad
$\dsize Z^i{}\!_\m = T^i{}\!_m h^m{}\!_\m $ are complex valued
functions of the real space-time coordinates $x^\a$. The complex
conjugate of  $Z^i{}\!_\m$ is just $\dsize \overline {Z^i{}\!_\m}
= \overline {T^i{}\!_m} h^m{}\!_\m $ because $h^i{}\!_\m$ remains
{\sl real}. It is easily seen from Eq.~(72) that Eq.~(73) yields
the same (real) metric as Eq.~(1), i.e., $\dsize g_{\m\n} = g_{ij}
h^i{}\!_\m h^j{}\!_\n $. Just as the real tetrad $h^i{}\!_\m$
provides a richer structure than $g_{\m\n}$ (a structure which
describes the gravitational and electroweak fields), the complex
tetrad $ Z^i{}\!_\m $ provides an even richer structure (a
structure which offers the possibility for describing the strong
interaction, while still describing gravity with the real metric
of general relativity). \subsubhead 4.1.1. Currents for Strong
Isospin and Hypercharge
\endsubsubhead
Working by analogy with Eq.~(40), we define $\frak C_i$ by
$$
\frak C_i = =- Z^{-1} \(Z Z_i{}\!^\n\)\!,_\n \quad  , \tag73
$$
where $Z$ is the determinant of $Z^i{}\!_\m$, and note that
$\dsize g^{ij}\overline {\frak C_i} \frak C_j $, is invariant not
only under real conservative coordinate transformations on Greek
indices, but also under complex conservative Lorentz
transformations on Latin indices, i.e., transformations $\dsize
Z^\mW{}\!_\m = L^\mW{}\!_i Z^i{}\!_\m $ which satisfy
$$
L^j{}\!_\mW  \(L^\mW{}\!_{i,j} - L^\mW{}\!_{j,i}\) =0 \quad \tag74
$$
and
$\dsize
g_{ij}= g_{\mW \nW} \overline {L^\mW{}\!_i} L^\nW{}\!_j
$
where
$
g_{ij}= g_{\mW \nW}=\text{diag}(-1,1,1,1).
$
For an infinitesimal complex Latin Lorentz transformation, one easily finds that\linebreak

\flushpar
$
L^i{}\!_\mW = \d^i_m + g^{ij} \f_{jm}
$
where
$
\f_{jm}
$
is anti-Hermitian.
The conservative condition, Eq.~(74), is satisfied if and only if
$
\f^i{}_{m,i}- \f^i{}_{i,m} = 0
$.
From the
$
\f^i\!{}_m
$
one can read off the generators for the transformation coefficients
$
L^i{}\!_\mW
$.

Field equations may be derived from a variational principle with Lagrangian
$\dsize
g^{ij}\overline {\frak C_i} \frak C_j
$.
The reality constraint on $g_{\m\n}$ is just
$$
g_{ij}\( \overline {Z^i{}\!_\m} Z^j{}\!_\n - {Z^i{}\!_\m} \overline {Z^j{}\!_\n}\)=0 \; .
$$
This constraint may be imposed by using Lagrange multipliers, and for the density $h$, we have
$\dsize
h=\sq{-g}= \sq{\overline Z Z}
$.
Thus, our variational principle is
$$
\d\I \[g^{ij}\overline {\frak C_i} \frak C_j  +
\L^{\m\n} g_{ij}\( \overline {Z^i{}\!_\m} Z^j{}\!_\n - {Z^i{}\!_\m} \overline {Z^j{}\!_\n}\)\] \sq{\overline Z Z} d^4x = 0 \tag 75
$$
where $Z^i{}\!_\m$,  $\overline {Z^i{}\!_\m}$ and $\L^{\m\n}$ are varied independently.
(Independent variation of $Z^i{}\!_\m$ and $\overline {Z^i{}\!_\m}$ is equivalent to varying the real and imaginary components of $Z^i{}\!_\m$ independently.)

After integration by parts, Noether's theorem gives (conserved) currents corresponding to strong isospin $I_3$ and hypercharge $Y$
$$\aligned
I_3 &= \overline {C^1} Z_1{}^\a - C^1 \overline {Z_1{}^\a}
         - \overline {C^2} Z_2{}^\a + C^2 \overline {Z_2{}^\a} \\
Y &= \overline {C^1} Z_1{}^\a - C^1 \overline {Z_1{}^\a}
         + \overline {C^2} Z_2{}^\a - C^2 \overline {Z_2{}^\a}
         -2\overline {C^3} Z_3{}^\a + 2C^3 \overline {Z_3{}^\a}
\endaligned \tag76
$$
It is clear that our discussion of the strong interaction is more speculative than the discussions in previous Sections. Much more work must be done before it may be possible to make a more definite claim.

The results presented in this paper indicate that this theory may lead to the fundamental
theory that unifies all the known forces. The theory contains no adjustable parameters. The standard model, by contrast, requires that many parameter values and symmetries must be "put in by hand."  The reason for this is that the standard model does not unify the electroweak and the strong interactions. And, of course gravity is not included in the standard model. In our theory, by contrast, gravity is present from the outset, and all forces are completely unified.
Indeed, our theory is constructed by analogy with general relativity, while the $U(1) \X SU(2)$ electroweak theory and the $SU(3)$ strong theory (the building blocks of the standard model)
are constructed by analogy with electromagnetism.

\subsubhead
4.2. Quantization of the theory.
\endsubsubhead
The theory thus far is at the classical level.  Before quantization via canonical
methods or path integrals, gauge constraints must be introduced to fix the gauge.  Type
I regions would require 16 gauge constraints, while none are required for Type II
regions. Alternatively one may introduce 16 fermionic ghost variables and their
conjugate momenta in Type I regions (Sundermeyer, 1982), (Henneaux and Teitelboim,
1992), (Weinberg, 1996).  These extra degrees of freedom act as negative degrees of
freedom which have the effect of fixing the gauge.

The quantized theory must be examined to determine whether it is
finite, or, at least, renormalizable and free of anomalies. There
are several reasons for believing that the quantized theory will
be either finite or renormalizable. First, Rosenfeld (1930) noted
certain advantages that tetrads present for the quantization of
gravity. Second, our Lagrangian $C^\mu C_\mu$ involves only first
derivatives of $h^i_{\text{ }\mu}$; whereas the Ricci scalar $R$,
the Lagrangian for gravitation alone, involves first and second
derivatives of $g_{\mu\nu}$.  Third, the conservation group is
much larger than the diffeomorphisms, and experience with gauge
theory suggests that larger groups offer more promise of
successful quantization. Fourth, we recall that the theory of weak
interactions alone was not renormalizable, but the theory became
renormalizable with the inclusion of electromagnetism.  This
provides hope that gravitation will become renormalizable with the
inclusion of the electroweak and/or strong interaction.

\subsubhead
4.3.  Fundamental geometrical issues.
\endsubsubhead
It is possible that certain geometric principles could
lead to a determination of coupling constants and masses.  The larger
symmetry of the conservation group suggests that the basic geometry is
not a space of points, but a space of paths.  Hence, we would
investigate connections between this theory and string theory. It appears possible that the path-space could provide a geometrical foundation for string theory. The need for such a foundation has been emphasized especially by Witten (1988), and Schwarz (1988) has noted that this foundation could be provided by a ``stringy space.''

\newpage

\head
{\bf REFERENCES}
\endhead

\flushpar Bade, W.~L., and H. Jehle (1953). ``An Introduction to Spinors,'' {\sl
Reviews\linebreak \hphantom{B} of Modern Physics,} {\bf 25}, 714.\newline Barut, A.~O.
(1980). {\sl Electrodynamics and Classical Theory of Fields and\linebreak \hphantom{B}
Particles, } 1st ed., Dover, New York, p. 11.\newline Bergmann, P.G. and Goldberg, I.
(1955). ''Dirac Bracket Transformation in\linebreak \hphantom{B} Phase Space,''
Physical Review {\bf 98}, 531.\newline Dirac, P.~A.~M. (1930). {\sl The Principles of
Quantum Mechanics,} Cambridge\linebreak \hphantom{B} University Press, Cambridge,
Preface to First Edition.\newline Dirac, P.A.M. (1964). {\sl Lectures on Quantum
Mechanics,} Academic Press,\linebreak \hphantom{B} New York.\newline Eddington, A.~E.,
(1924). {\sl The Mathematical Theory of Relativity,} 2nd~ed,\linebreak \hphantom{B}
Cambridge University Press, Cambridge, p.~222.\newline Einstein, A. (1928A).
``Riemanngeometrie mit Aufrechterhaltung des\linebreak \hphantom{B} Begriffes des
Fernparallelismus,'' {\sl Preussischen Akademie der Wissenshaf-\linebreak
\hphantom{Bt}ten, Phys.-math. Klasse, Sitzungsberichte}\/ {\bf  1928}, 217.\newline
Einstein, A. (1928B). ``Neue M\"oglichkeit f\"ur eine einheitliche Feltheorie
von\linebreak \hphantom{B} Gravitation und Elektrizit\"at,'' {\sl Preussischen Akademie
der Wissenshaf-\linebreak \hphantom{Bt}ten, Phys.-math. Klasse, Sitzungsberichte}\/
{\bf  1928}, 224.\newline Einstein, A. (1949), in {\sl Albert Einstein:
Philosopher-Scientist,} edited by\linebreak \hphantom{B} P. A. Schilpp, Harper \&
Brothers, New York, Vol. I, p. 89.\newline Eisenhart, L.~P. (1925). {\sl Riemannian
Geometry,} Princeton University Press,\linebreak \hphantom{B} Princeton, p~97.\newline
Finkelstein, D. (1981). Private communication.\newline Green, E.~L. (1991). Reported in
Pandres (1995).\newline Green, E.~L. (1997). Reported in Pandres (1999).\newline
Henneaux, M. and Teitelboim, C. (1992). {\sl Quantization of Gauge Systems,}\linebreak
\hphantom{B} Princeton University Press, Princeton.\newline Loos, H.~G. (1963). ``Spin
Connection in General Relativity,'' {\sl Annals of\linebreak \hphantom{B} Physics,}
{\bf 25}, 91.\newline Misner, C. (1957). ''Feynman Quantization of General
Relativity,'' Reviews\linebreak \hphantom{B} of Modern Physics {\bf 29},
497-509.\newline Moriyasu, K. (1983). {\sl An Elementary Primer for Gauge Theory,}
World\linebreak \hphantom{B} Scientific, Singapore, p.~110.\newline Nakahara, M.
(1990). {\sl Geometry, Topology and Physics,} Adam Hilger, New\linebreak \hphantom{B}
York, p.~344.\newline Pandres, D., Jr. (1962). ``On Forces and Interactions between
Fields,''\linebreak \hphantom{B}{\sl Journal of Mathematical Physics,}{\bf 3},
602.\newline Pandres, D., Jr. (1981). ``Quantum Unified Field Theory from
Enlarged\linebreak \hphantom{B} Coordinate Transformation Group,'' {\sl Physical Review
D,} {\bf 24}, 1499.\newline Pandres, D., Jr. (1984A). ``Quantum Unified Field Theory
from Enlarged\linebreak \hphantom{B} Coordinate Transformation Group. II,'' {\sl
Physical Review D,} {\bf 30}, 317.\newline Pandres, D., Jr. (1984B). ``Quantum Geometry
from Coordinate Transf-\linebreak \hphantom{B}ormations Relating Quantum Observers,''
{\sl International Journal of Theo-\linebreak \hphantom{B}retical Physics,} {\bf 23},
839.
\flushpar Pandres, D., Jr. (1995). ``Unified Gravitational and Yang-Mills
Fields,''\linebreak \hphantom{B} {\sl International Journal of Theoretical Physics,}
{\bf 34}, 733.\newline Pandres, D., Jr. (1998). ``Gravitational and Electroweak
Interactions,''\linebreak \hphantom{B} {\sl International Journal of Theoretical
Physics,} {\bf 37}, 827-839.\newline Pandres, D., Jr. (1999). ``Gravitational and
Electroweak Unification,''\linebreak \hphantom{B} {\sl International Journal of
Theoretical Physics,} {\bf 38}, 1783-1805.
\newline Rosenfeld, I. (1930). Zur Quantelung der Wellen felder, {\sl Annalen der
Physik}, {\bf 5}, 113.
\newline Salam, A. (1968). ``Weak and
Electromagnetic Interactions,'' {\sl Proceedings\linebreak \hphantom{B} of the 8th
Nobel Symposium on Elementary Particle Theory,} edited by\linebreak \hphantom{B} N.
Svartholm, Almquist Forlag, Stockholm, p.~367.\newline Schouten, J.~A. (1954). {\sl
Ricci-Calculus, 2nd Ed.} North-Holland, Amsterdam,\linebreak \hphantom{B}
p.~99ff.\newline Schr\"odinger, E. (1960). {\sl Space-Time Structure,} Cambridge
University Press,\linebreak \hphantom{B} Cambridge, p.~97,~99.\newline Schwarz, J.
(1988). In {\sl Superstrings: A Theory of Everything ?,} Edited by\linebreak
\hphantom{B} P.~C.~W. Davis and J. Brown, Cambridge University Press,
Cambridge,\linebreak \hphantom{B} p.~70.\newline Sundermeyer, K. (1982). {\sl
Constrained Dynamics,} Springer-Verlag, Berlin.\newline Synge, J.~L. (1960). {\sl
Relativity: The General Theory,} North-Holland,\linebreak \hphantom{B} Amsterdam,
p.~14, 357.\newline Weber, J. (1961). {\sl General Relativity and Gravitational Waves,}
Interscience,\linebreak \hphantom{B} New York, p.~147.
\flushpar Weinberg, S. (1967).
``A Model of Leptons,'' {\sl Physical Review Letters,} {\bf 19},\linebreak \hphantom{B}
1264.\newline Weinberg, S. (1995). {\sl The Quantum Theory of Fields, Vol.I},
Cambridge\linebreak \hphantom{B} University Press, Cambridge.\newline Weinberg, S.
(1996). {\sl The Quantum Theory of Fields, Vol.II}, Cambridge\linebreak \hphantom{B}
University Press, Cambridge.\newline Weitzenb\"ock, R. (1928). ``Differentialivarianten
in der Einsteinschen\linebreak \hphantom{B} Theorie de Fernparallelismus,'' {\sl
Preussischen Akademie der Wissenshaften,\linebreak \hphantom{B} Phys.-math.
Klasse,Sitzungs-berichte} {\bf  1928}, 466.\newline Witten, E. (1988). In {\sl
Superstrings: A Theory of Everything ?,} Edited by\linebreak \hphantom{B} P.~C.~W.
Davis and J. Brown, Cambridge University Press, Cambridge,\linebreak \hphantom{B}
p.~90.\newline

\enddocument